\documentclass[epj]{webofc}
\usepackage[varg]{txfonts}   
\woctitle{MESON2014 - the 13$^\textrm{th}$ International Workshop on Meson Production, Properties and Interaction}
\begin{document}
\selectlanguage{english}
\title{Search for baryonium and the physics at FAIR}

\author{S. Wycech \inst{1}\fnsep\thanks{\email{wycech@fuw.edu.pl}} \and
        J-P. Dedonder \inst{2} \and
        B. Loiseau \inst{2}
}

\institute{National Centre  for Nuclear Studies, Warsaw, Poland 
\and
Sorbonne Universit\'es, Universit\'e Pierre et Marie Curie, Sorbonne Paris Cit\'e, Universit\'e Paris Diderot,
et IN2P3-CNRS, UMR 7585, LPNHE, 4 place Jussieu,
75252 Paris, France
          }

\abstract{%
The existence of nucleon-antinucleon
quasi-bound state is indicated  in two decay modes of the
$J/\psi$ meson studied by BES. We discuss  an explanation in a fairly traditional
Paris-potential model of the $N\overline{N}$ interactions.  A  broad, 
$S$-wave and a narrow $P$-wave quasi-bound states    are predicted by this model. 
Some existing  experimental evidence and possible    verifications in the future are indicated. 
}
\maketitle
\section{Introduction}
\label{s-1}

 Baryonium understood as a nucleon-antinucleon quasi-bound was searched for  at CERN in the days of  LEAR. Nothing has been  found,  
 but broad states or states close to the  threshold  were not excluded~\cite{BER82, ADI86}. One possible reason of the failure is  the heavy
background due to annihilation processes. Another is the large number of allowed partial vaves.   A convincing detection requires  selective  
experiments, and the  first such  experiment was performed   by the BES Collaboration~\cite{BAI03} who studied the decay  
\begin{equation}
\label{1}
 J/\psi \rightarrow \gamma p\overline{p}.
\end{equation}
The  selectivity in this reaction  is due to the definite initial state and CP invariance. As shown in figure~\ref{fig-3}  
the measured  spectrum  of  $p\overline{p}$ invariant mass displays two prominent maxima named  initially $ X(1859)$  and $ X(2170)$. 
The first one, close to the threshold, is apparently due to strong attraction in the final iso-singlet  spin-singlet $ ^{11}S_0$ wave which is 
one of three partial waves allowed in reaction~(\ref{1}). Indeed, calculations of final state  interactions (FSI) performed   with  
Paris~\cite{LOI05} and J\"ulich~\cite{HAI06}
potential indicate a strong threshold enhancement.  The basic difference is that with the Paris potential  the threshold  peak is due to a bound state 
and  with the J\"ulich potential it results from a virtual state. To distinguish between these two possibilities, one  needs to look directly below the  $p\overline{p}$ threshold. Several methods to do this are discussed in section~\ref{sec-2}.

\subsection{Final state interactions in $J/\psi$ decays}
\label{ss-1}
\begin{figure}[ht]
\centering
\sidecaption
\includegraphics[width=4cm,clip]{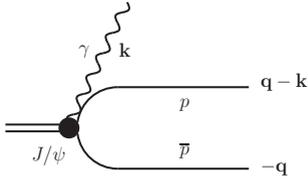}
\caption{The photon is  emitted either from $J/\psi$ or during the hadronisation stage of the process and the final baryons are formed 
in the $S$ wave}
\label{fig-1}       
\end{figure}

\begin{figure}[ht]
\centering
\sidecaption
\includegraphics[width=5cm]{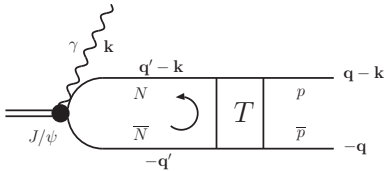}
\caption{Final state interactions described by half-off shell  $T$ matrix 
generated by the  Paris potential}
\label{fig-2}       
\end{figure}

In this note  we report an extension of the FSI calculations  of Ref.~\cite{LOI05} which  now covers  the whole photon  spectrum. The basic 
assumption of this approach  (also that in  Ref.~\cite{HAI06}) is that the photon is emitted before the baryons are formed. The two  related processes 
are plotted in diagrams~\ref{fig-1},  \ref{fig-2} and the FSI is calculated in terms of half-off shell $T$ matrix 
generated by the Paris potential~\cite{LAC09}.
This approach allows to calculate the spectrum but not the absolute decay rate. One free parameter, the radius $R$ (=0.28~fm)
of a  Gaussian source function is used to describe  the creation of a $\gamma p\overline{p}$  state (see figure~\ref{fig-1}). However, in order to reproduce  in a better way both maxima 
$ X(1859) $ and $X(2170)$, it turns out profitable to assume the radius to be weakly  dependent  on the photon energy.
It was found to  change from $0.28$~fm at maximal  $k\sim1.2$ GeV
to $0.39$~fm at  $k=0$.   

\begin{figure}[ht]
\centering
\sidecaption
\includegraphics[width=5cm, angle=90]{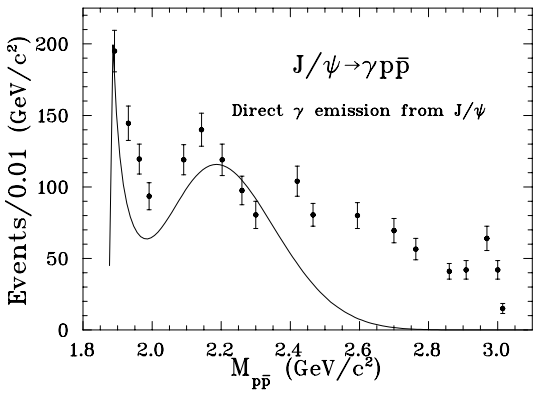}
\caption{The  $p \overline{p}$  invariant mass spectrum  obtained under the assumption that photon is emitted  before the baryons are formed. 
The missing strength at large $p\overline{p}$ invariant mass, $M_{p\overline{p}}$, comes from the photon radiated by final hadrons~\cite{DED14}
 }
\label{fig-3}       
\end{figure}

\begin{figure}[ht]
\centering
\sidecaption
\includegraphics[width=5cm, angle=90]{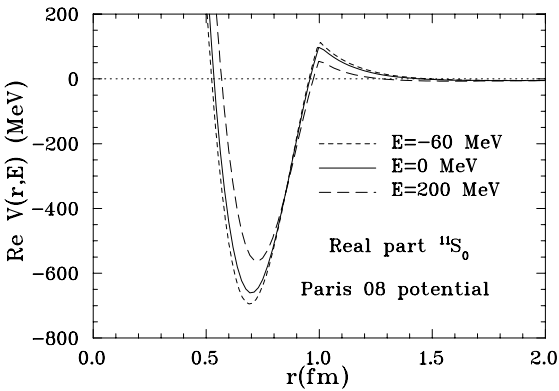}
\caption{The Paris $N \overline{N}$ real  potential in the $ ^{11}S_0$ wave. It generates a $50$~MeV broad quasi-bound state 
at~$ \sim 5 $~MeV binding. The well and   barrier structure generate the  shape resonance visible in the spectrum 
in figure \ref{fig-3}  at~$ 2170 $ MeV}
\label{fig-4}       
\end{figure}
Inspection of figure~\ref{fig-3} shows that both states may be reproduced by the  potential   $N \overline{N}$ interactions related at 
large distances via G-parity transformation to the $N N$ interactions. However,  the proper description of both peaks involves 
distant extrapolation of the $T_{N\overline{N}}$ matrices off energy shell, which  corresponds to  very short ranged interactions. 
This figure shows also that a 
sizable portion of the spectrum is missing. The missing  part  comes from  photon emissions by final baryons: $N\overline{N}$ and exchange currents
\cite{DED14}. Similar    final state emission model  offers
a consistent description of the rate and the spectra in cases of $ \pi,\omega,\phi$ mesons with a constant  $R=.28$~fm radius. 
 This  knowledge of  initial formation radius  allows to calculate  inverse
reactions leading to formation of $ J/\psi$ and a meson in nuclear collisions of $\overline{p}$~\cite{DED14}.

\section{Looking below the $N \overline{N}$ threshold}
\label{sec-2}
Testing  the  subthreshold amplitudes  may be realised in few body systems in particular in light antiprotonic atoms or at 
extreme nuclear peripheries.
 In these conditions nucleons are bound and the effective subthreshold energies are  composed of binding energies and recoil  of the $N\overline{p}$ 
 pair  with respect to the rest of the system. For valence nucleons the   $ - E_{binding} - E_{recoil} $ may reach down to - 40~MeV. 
 Here, we want to point out one experiment that might be of interest at FAIR/FLAIR. 
Table~\ref{tab:radio} shows the ratios of antiproton capture rates  on neutrons and protons  $ C(n \overline{p})/ C(p \overline{p})$ bound to nuclear peripheries. These reflect the ratios of neutron and proton  densities. The second and third columns indicate such ratios extracted from   widths of two    antiprotonic 
atomic  levels the "lower" and the "upper" one. These widths are determined at  nuclear densities   $ \sim 10 \% $ and $ \sim 5\%$ of the central density $\rho_0$. The last  column is obtained
with radiochemical studies of final nuclei with  one neutron or one proton removed in the annihilation reaction~\cite{LUB97}. 
The latter process is localised
at densities  $\rho \sim 10^{-3} \rho_0$.  In standard nuclei shown in the upper part of the table the ratios $ C(n \overline{p})/ C(p \overline{p}) \sim \rho_n /\rho_p $
increase at nuclear peripheries. However, in some nuclei characterised by a  small proton binding, indicated  in the lower part of the table,  the ratio $ C(n \overline{p})/ C(p \overline{p})$ suddenly drops at extreme nuclear peripheries. That effect cannot be explained by the nuclear structure alone  and we attribute it to the
existence  of a narrow
bound state in the $N \overline{N}$ system. Such a narrow state is in fact predicted by the Paris potential  in the  $^{33}P_1$  wave, see table~\ref{tab-1}.

\begin{table}[ht]
\centering
\caption{Ratios of  $ N(n \overline{p}) $ and $ N(p \overline{p}) $ capture
rates from atomic states. The last  column  shows experimental
numbers from radiochemical experiments. Other columns (see text) give ratios
calculated with optical potential and plausible nuclear
densities  based on experimental results
from Ref.~\cite{LUB97}}
 \label{tab:radio}
  \begin{tabular}{llll}
  \hline
atom & lower & upper& radiochemistry  \\\hline
 $^{96}$Zr &0.95(9)&1.53(29)&2.6(3) \\
$^{124}$Sn &1.79(10)&2.44(39)&5.0(6) \\ \hline
$^{106}$Cd &1.64(80)&2.10(80)&0.5(1) \\
$^{112}$Sn &1.90(13)&2.43(49)&0.79(14)\\
\end{tabular}
\end{table}
 
Another way to look below threshold is   opened by the BES collaboration experiment~\cite{ABL05}
detecting  radiative $J/\psi$ decays
\begin{equation}
\label{2} 
 \ J/\psi \rightarrow  \gamma   \pi^+ \pi^- \eta' 
\end{equation}
and observation that the mesons are correlated to another state named $ X(1835)$ with quantum numbers consistent with the $ X(1859)$ state.  Within the Paris-potential interpretation, the $ X(1835)$  is due to the interference of the baryonium  $^{11}S_0$ state, seen above the threshold, with a 
background  formation amplitude~\cite{DED09}. 

\begin{table}[ht]
\centering
\caption{Binding energies  in MeV of the close to
threshold quasi-bound states in  the Paris potential ~\cite{LAC09} }
\label{tab-1}       
\begin{tabular}{ll}
\hline
$^{2T+1\ 2S+1}L_J$ &  $ E - i \Gamma/2$   \\\hline
$^{11}S_0$ & -4.8-i26  \\
$^{33}P_1$ & -4.5-i9.0  \\\hline
\end{tabular}
\end{table}

\subsection{Experiments suggested }
Following the indications from BES experiments the  baryonia should be searched in the region of $0 - 60$  MeV below  the $N \overline{N}$ 
threshold. It would be advisable to repeat two old experiments at  different energies, possibly with polarized particles :

$\bullet$ Search for narrow signals in the $\gamma$-spectrum from $p \overline{p}$ annihilation was performed  at rest \cite{ADI86},
The signals (in the region that we  expect them now to exist) were  covered by heavy background due to $\pi^0$ decays and 
$\pi^-p \rightarrow \gamma n $. It would be better to perform this experiment with few hundred MeV antiprotons which would 
shift the expected signal away from the region of  high background.

$\bullet $  The  $ \overline{p} d \rightarrow n  X $ experiment \cite{BER82} was performed at $ 1.3 $ GeV/c. This gives rather  small 
chance of $p\overline{p}$ coupling in the statistically insignificant $ ^{11}S_0$ wave. Lower energies would be better. 

New instructive experiments that possibly could be performed in the coming years are 
 
$\circ$  Fine structure splitting in light antiprotonic atoms $^1H,^2H,^3H,^3He,^4He$ would allow to trace energy dependence 
of the selected $ \overline{p}N$ amplitudes in the subthreshold region down to $ \sim - 40 $ MeV.

$\circ$  Studies of mesons emitted  from annihilations of $ \overline{p} $  at nuclear peripheries. In particular nuclei with closed shells 
 with one loosely bound valence nucleon could be profitable. In the latter case the baryonium  signal would be separated from 
 a complicated background due to  other annihilation channels.

\begin{acknowledgement}
This work has been partially supported by a grant from the French-Polish exchange program COPIN/CNRS-
IN2P3, collaboration 05-115.
SW was also supported by Narodowe Centrum Nauki grant 2011/03/B/ST2/00270

\end{acknowledgement}
%
%
%

\end{document}